\documentclass[aps,prl,twocolumn,superscriptaddress]{revtex4}
\usepackage{graphicx}

\begin{document}



\title{Aluminum arsenide cleaved-edge overgrown quantum wires}


\author{J.~Moser}
\email{moser@wsi.tu-muenchen.de}
\affiliation{Walter Schottky Institut, Technische Universit\"at
M\"unchen, D-85748 Garching, Germany\\}

\author{T.~Zibold}
\affiliation{Walter Schottky Institut, Technische Universit\"at
M\"unchen, D-85748 Garching, Germany\\}

\author{S.~Roddaro}
\affiliation{Scuola Normale Superiore, via della Faggiola, I-56126 Pisa, Italy\\}

\author{D.~Schuh}
\affiliation{Walter Schottky Institut, Technische Universit\"at
M\"unchen, D-85748 Garching, Germany\\}

\author{M.~Bichler}
\affiliation{Walter Schottky Institut, Technische Universit\"at
M\"unchen, D-85748 Garching, Germany\\}

\author{V.~Pellegrini}
\affiliation{Scuola Normale Superiore, via della Faggiola, I-56126 Pisa, Italy\\}

\author{G.~Abstreiter}
\affiliation{Walter Schottky Institut, Technische Universit\"at
M\"unchen, D-85748 Garching, Germany\\}

\author{M.~Grayson}
\affiliation{Walter Schottky Institut, Technische Universit\"at
M\"unchen, D-85748 Garching, Germany\\}


\date{28 Feb 2005}


\begin{abstract}

We report conductance measurements in quantum wires made of aluminum arsenide,
a heavy-mass, multi-valley one-dimensional (1D) system. Zero-bias conductance steps
are observed as the electron density in the wire is lowered, with additional steps observable
upon applying a finite dc bias.  We attribute these steps to depopulation of successive 
1D subbands.  The quantum conductance is substantially reduced with respect to the
anticipated value for a spin- and valley-degenerate 1D system. This reduction is consistent with disorder-induced, intra-wire backscattering which suppresses the transmission of 1D modes. 
Calculations are presented to demonstrate the role of strain in the 1D states of this cleaved-edge structure.

\end{abstract}


\maketitle



One-dimensional systems of heavy electrons (effective mass $m^* = m_{0}$, where $m_{0}$ is the free electron mass) are promising candidates for the study of electronic correlations. However owing to an inherent low mobility $\mu \sim 1/m^{*}$ and fabrication challenges, few realizations are reported in the literature.  Experiments in Si \cite{oda,klapwijk} and Si/SiGe \cite{abstreiter1,wieser} quantum point contacts (QPC) have only focused on transport in (100)-plane structures with light isotropic mass $m^{*}=0.19 m_{0}$. A conductance step height $G_{0}=g_s g_v e^2/h=4~e^{2}/h$ was found in the ballistic regime \cite{abstreiter1}, accounting for both spin $g_s = 2$ and valley $g_v = 2$ degeneracy, and $G_{0}=2~e^{2}/h$ was observed in those systems in the presence of disorder \cite{abstreiter1}. Aluminum arsenide (AlAs) is an alternate heavy mass system with degenerate valleys and anisotropic mass. In AlAs the constant energy surfaces are ellipses centered at the $X_i$ points $(i = x, y, z)$ at the Brillouin zone edge, characterized by a heavy longitudinal mass $m^*_H = 1.1~m_{0}$ and light transverse mass $m^*_L = 0.19~m_{0}$ \cite{adachi}. The energy minima at these $X$ points are highly sensitive to strain \cite{strain}.

In this paper we report conductance measurements of a quantum wire fabricated at the edge of an aluminum arsenide (AlAs) 2D-electron gas (2DEG) using cleaved-edge overgrowth (CEO). In absence of interactions a valley degeneracy $g_v = 2$ is predicted in the wire. Two conductance steps are observed at low electron density, with step height $G_{0} \sim 0.44~e^{2}/h$. Applying a finite dc bias across the wire reveals cleaner steps of the same height. We discuss the role of strain, disorder and interactions in relation to the observed 1D conductance.

AlAs wires are fabricated with the same CEO technique employed for ballistic GaAs wires \cite{yacoby_solidstatecom}. Our samples contain a modulation-doped AlAs quantum well \cite{Shayegan} flanked by two ${\rm Al}_x{\rm Ga}_{1-x}{\rm As}$ dielectric barriers $(x = 0.45)$
and grown onto a [001]-oriented GaAs substrate by molecular beam epitaxy (MBE). The 150~{\AA} wide quantum well resides 4000~{\AA} below the surface with density $n=5.6 \times 10^{11}$~cm$^{-2}$ and mobility $\mu=$~55,000~cm$^{2}$/Vs at $T=340$~mK.  A 1~$\mu$m-wide tantalum gate is patterned on top of the heterostructure (Fig.~1, inset).  Samples are cleaved {\it{in-situ}}, and the exposed (110) facet is overgrown with a modulation-doping sequence. Applying a negative bias to the top gate depletes the 2DEG, leaving behind a 1D accumulation wire that is electrically contacted with the ungated 2DEG regions on either side.

Fig.~1, right displays the wire conductance $G$ measured as a function of gate bias $V_{g}$ at fixed temperature $T=20$ and 340~mK, following illumination from the backside at $T \simeq{10}$~K with an infrared LED. The wire was biased with an excitation voltage $V_{ac}=10\mu V<2\pi k_{B}T/e$, where $e$ is the electron charge. Each $G(V_{g})$ trace corresponds to a different cooldown and illumination, and therefore to a slightly different electron density. Reproducible features appear in a 1:2:3 ratio along the $G-$axis as steps close to $G=0.44$ and $0.88~e^{2}/h$, and as a change in average slope $dG(V_{g})/dV_{g}$ at $G\simeq 1.32~e^{2}/h$, indicating a quantum conductance step $G_{0}=0.44~e^{2}/h$. A conductance quantization of $0.44\pm 0.05~e^{2}/h$ was similarly observed in other samples. Below $G_{0}$ fluctuations are visible in the 340~mK trace, reminiscent of what is observed in disordered CEO GaAs wires \cite{yacoby_prl96,auslaender_prl00,rother_thesis}. Step-like features are observed below $T \simeq 1$~K.

More conductance steps appear with the application of a dc source-drain bias $V$(Fig.~1, left). The two lowest steps labeled 1 and 2 show stable step values comparable to those at $V=0$, and other features labeled 3 and upwards show a step-like structure whose conductance values are increasingly dependent on the bias voltage. Comparing to the zero bias traces, one sees that the position of the first zero bias step has shifted from $V_g = -2.45$~V to $-2.48$~V. The clear step structure both at finite $V$ and $V=0$ is the strongest evidence of the 1D nature of our wires.

Strain is important in AlAs \cite{shkolnikov}, so we pause to clarify its influence on the subband energies and valley degeneracy $g_v$ in this structure before continuing with the analysis.   In a local strain field, the strain components that control $X$-valley splitting are $\epsilon_{xx}$, $\epsilon_{yy}$ and $\epsilon_{zz}$ for the $X_x$, $X_y$ and $X_z$ points, respectively. In 2D AlAs systems, strain splits the degeneracy of the valleys from a mean value $E_0$ by an amount $\Delta$ \cite{strain}, which is then compensated by the quantum confinement $W$ in the $z$ direction:

\begin{eqnarray}
E_{x} = E_{y} = E_{xy} & = & E_0 + \frac{\hbar^2 \pi ^2}{2 m^*_L W^2} - \frac{\Delta}{2} \label{2Dsplitting}
\\
E_{z} & = & E_0 + \frac{\hbar^2 \pi ^2}{2 m^*_H W^2} + \frac{\Delta}{2} \nonumber
\end{eqnarray}

\noindent 
We simulated \cite{NextNANO} these strain-induced splittings with matrix calculations of the 3-dimensional strain tensor $\epsilon$ \cite{herring} for 2D quantum well systems and verified the experimentally demonstrated degeneracy crossover  $E_{xy} = E_{z}$ at $W \sim 60$~\AA\cite{strain}. For the wider quantum wells of this work, the strain term dominates Eq.~\ref{2Dsplitting} and $E_{xy} < E_{z}$, yielding dual valley degeneracy.

\begin{figure}[b]
\includegraphics[width=8.5cm]{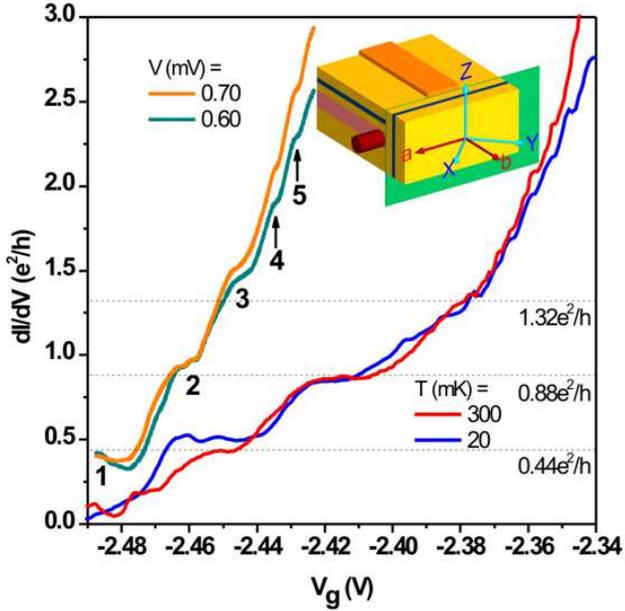}
\caption{(color online) Right: conductance $G$ of an AlAs quantum wire as a function of gate bias $V_{g}$, at bath temperatures $T=$20~mK (blue) and 300~mK (red) for two different cooldowns. Left: the population of higher 1D subbands (3, 4, 5) can be observed at $T=20$~mK upon applying a finite source-drain dc bias $V$. Inset: schematic of the sample. The protruding red cylinder represents the wire at the edge of the AlAs 2DEG. Yellow: AlGaAs; blue: doping layers; green: cleavage plane; orange: top gate.}
\label{fig1}
\end{figure}

For the CEO structure, we extended these strain calculations to determine the various spatially dependent $X$-band energies near the cleave plane. With $\hat{x}, \hat{y}, \hat{z} = \left[100\right], \left[010\right], \left[001\right]$ as unit vectors along the crystal axes, we define two other useful unit vectors $\hat{a} = (\hat{x}-\hat{y})/\sqrt{2} = \left[1\bar{1}0\right]$ parallel to the wire, and $\hat{b} = (\hat{x} + \hat{y})/\sqrt{2} = \left[110\right]$ normal to the cleave-plane. The spatial dependence of strain in the $(\hat{b},\hat{z})$ plane perpendicular to the wire is depicted in the color plot of Fig.~2 with red showing maximum compressive strain in the $\epsilon_{xx} = \epsilon_{yy}$ components in Fig.~2a, and red showing maximum tensile strain in $\epsilon_{zz}$ in Fig.~2b. The symmetry of the structure requires $\epsilon_{xx}=\epsilon_{yy}$, so the $X_x$ and $X_y$ valleys are shifted by the same amount and the valley degeneracy $g_v = 2$ in the lowest bands of the 2D well is preserved even in the wire region.

\begin{figure}[t]
\includegraphics[width=8.5cm]{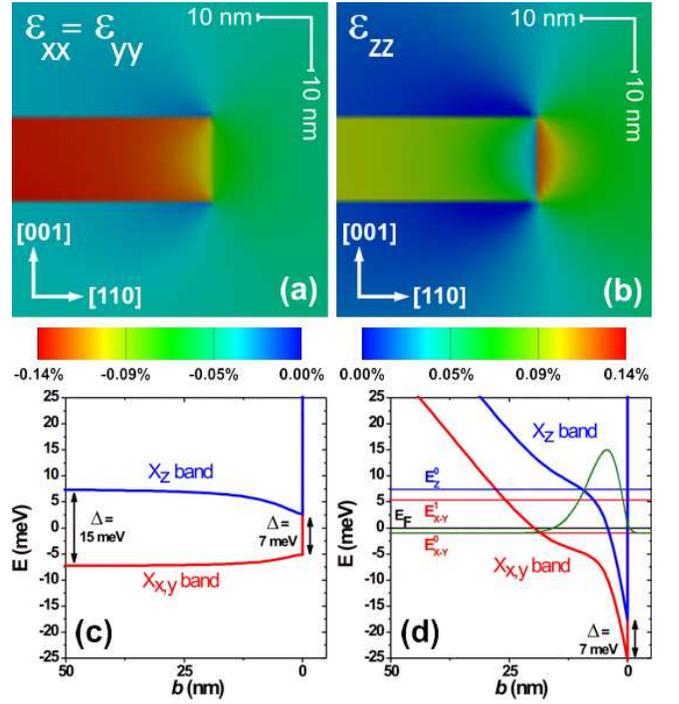}
\caption{(color online) Compressive (a) and tensile (b) strain components in the $(\hat{b},\hat{z})$ plane. (c)-(d) $X_{xy}$ and $X_{z}$ band structure in the $\hat{b}$ direction within the effective-mass approximation (distance is measured from the cleavage plane). In (c) only strain is included. The energy $E$ is offset by an arbitrary amount. In (d) both strain and electrostatics are computed. $E_{i}^{0,1}$ is the energy of the first, second $X_{i}$ 1D subband. $E_{F}=0$ is the Fermi level. The probability density for the first $X_{xy}$ 1D state is indicated.}
\label{fig2}
\end{figure}

Fig.~\ref{fig2}c shows the effect of strain alone (no doping) on the $X_{z}$ and $X_{xy}$ bands in the $\hat{b}$ direction. Far from the cleavage plane, the valley splitting of the band edges reaches 15~meV, however since both compressive $\epsilon_{xx} = \epsilon_{yy}$ and tensile $\epsilon_{zz}$ strain components decrease in magnitude at the cleave plane, the $X_{xy}$ and $X_{z}$ bands approach each other reaching a 7~meV separation at the cleave-plane. In Fig.~\ref{fig2}d the band calculation includes doping in the 2DEG and the overgrown structure; the 2DEG has been depleted underneath the gate. The energies of the first and second $X_{xy}$ 1D subbands are still below the energy of the first $X_{z}$ 1D subband.

Finally, we note the effect of the cleave-plane geometry on the 1D band mass. In units of the free electron mass, the scalar effective mass $m_{\hat{r}}$ projected along a given direction $\hat{r}$ follows $1/m^*_{\hat{r}} = \sum_{jk} \hat{r}_j (m^*)^{-1}_{jk} \hat{r}_k $ with $(m^*)^{-1}$ the inverse mass tensor, $m^*_{11}=m^*_H, m^*_{22} = m^*_{33}=m^*_L,$ and $m^*_{jk} = 0$ for $j\neq k$. The result for unit vectors along the wire $\hat{a}$ and perpendicular $\hat{b}$ are $m^*_{\hat{a}}=m^*_{\hat{b}}=0.33~m_{0}$, almost a factor of 2 larger than $m^*$ in (100) Si and a factor of 5 larger than $m^{*}$ in GaAs wires. Using a calculated energy spacing between the first and second 1D subbands $\delta\simeq 6$~meV and a Fermi velocity $v_{F}$ corresponding to the filling of the first subband, one sees that such a large $m^{*}$ renders temperature irrelevant above $\sim 1$~K as the thermal length $L_{th}=hv_{F}/(2k_{B}T)$ exceeds the wire length $L=1 \mu$m.

We now discuss the experimentally measured step height $G_{0} = 0.44 e^{2}/h$, having clarified the effect of the strain and 1D confinement on the wire degeneracy. Within the Landauer-B\"{u}ttiker formalism, the 1D conductance quantum reads $G_{0}=g_{s}g_{v}\xi e^{2}/h$, where $g_{s} = g_{v}=2$ denotes the spin and valley degeneracy and $\xi$ is the transmission factor, so that $G_{0}=4e^{2}/h$ in the ballistic regime, almost a factor of 10 larger than measured.

Although a dominant factor in reducing GaAs CEO wire steps is 1D-2D non-adiabatic coupling, in these AlAs wires it cannot account for such a large reduction. Instead, the simplest explanation for the suppression of $G_{0}$ is a reduction of the transmission factor $\xi$ induced by the random distribution of backscattering centers. Those centers may be related to interface roughness or remote ionized dopants. Indeed, the absence of 'half-plateaus' \cite{kouwenhoven} in the differential conductance $dI/dV$ as a function of $V_{g}$ (see Fig.~1) may indicate that $V$ drops at several scattering events along the wire. Disordered GaAs CEO wires of length $L>l_{B}^{wire}$, where $l_{B}^{wire}$ is the backscattering length in the wire, also exhibit a largely reduced conductance: $G_{0}\simeq 0.4\times 2e^{2}/h$ \cite{auslaender_prl00,rother_thesis}. Following Ref.~\cite{bagwell} we propose that under bias the transmission coefficient $\xi$ across the scattering potential landscape of the wire is averaged over a potential window of width $eV$, thereby averaging out conductance fluctuations and making higher 1D subband features apparent.

The assumption of a 4-fold degeneracy, however, is not straightforward in such 1D systems.  The reduced conductance step might be partly due to a degeneracy lifting in the wire. For example, the so-called 0.7-structure was suggested to be a spontaneous spin-polarization at zero external magnetic field in GaAs QPC \cite{thomas,abi}. In addition, $G_{0}=1e^{2}/h$ has recently been reported in single-walled carbon nanotubes \cite{marcus}, another two-band 1D material which in the absence of interactions should contain $g_{v}g_{s}=4$ conductance modes. Such degeneracy-breaking in $g_{v}$ or $g_{s}$ are proposed to occur at sufficiently low densities above critical values of the Wigner-Seitz parameter $r_{S}$ \cite{gold2}. If we fill the first subband to 6 meV then the effective Bohr radius $a^{*}=20$~{\AA} (using $m^{*}=0.33~m_{0}$) yields $ r_{S} = \pi /(2k_{F} g_{v} g_{s} a^{*}) \simeq 0.8$, where $k_{F}$ denotes the Fermi wavevector in the wire, and $g_{v}=g_{s}=2$.  AlAs wires therefore reach an interesting regime for 1D systems, in that the second subband begins to become occupied while the first subband is still near the highly interacting $r_S \sim 1$ limit.  Such strong interactions may play a role in breaking the 4-fold spin/valley degeneracy, making this system of interest for future theoretical and experimental study.


In summary, we have measured conductance quantization in AlAs quantum wires with $G_{0}\simeq 0.44~e^{2}/h$. Strain at the cleaved-edge should preserve the valley degeneracy $g_{v}=2$ within the wire. $G_{0}$ is smaller than the single-channel conductance quantum, and is substantially reduced with respect to the conductance quantum $4~e^{2}/h$ anticipated for spin- and valley-degenerate ballistic wires. This reduction suggests a backscattering-induced suppression of the transmission coefficient. A narrower gate and an enhanced 2DEG mobility may help access the ballistic regime in AlAs quantum wires.     

$\\$
J.M and M.G would like to thank  C.~Chamon, K. Kempa, and T. Giamarchi for helpful discussions.
J.M. gratefully acknowledges support from the COLLECT EC-Research Training Network, HPRN-CT-2002-00291.





\end{document}